\documentclass[useAMS,usenatbib]{mn2e}

\include{epsf}

\def\gsim{\lower.73ex\hbox{$\sim$}\llap{\raise.4ex\hbox{$>$}}$\,$}
\def\lsim{\lower.73ex\hbox{$\sim$}\llap{\raise.4ex\hbox{$<$}}$\,$}
\def\mpc{$\,h^{-1}\,$Mpc}
\def\%{~per~cent}

\title[The 2MASS galaxy angular power spectrum]
{The 2MASS galaxy angular power spectrum: Probing the galaxy distribution to Gigaparsec scales}
\author[W.J. Frith, P.J. Outram \& T. Shanks]
{W.J. Frith\thanks{E-mail:w.j.frith@durham.ac.uk}, 
P.J. Outram \& T. Shanks\\
Dept. of Physics, Univ. of Durham, South Road, Durham DH1 3LE, UK}
\begin{document}

\date{Accepted 2005. Received 2005; in original form 2005 }

\pagerange{\pageref{firstpage}--\pageref{lastpage}} \pubyear{2005}

\maketitle

\label{firstpage}

\begin{abstract}
We present an angular power spectrum analysis of the 2 Micron All Sky Survey (2MASS) full release extended source catalogue. 
The main sample used includes 518~576 galaxies below an extinction-corrected magnitude of $K_s=$13.5 and limited to 
$|b|>20^{\circ}$. The power spectrum results provide an estimate of the galaxy density fluctuations at extremely large scales, $r$\lsim 1000 
\mpc. We compare this with mock predictions constructed from the $\Lambda$CDM Hubble Volume mock catalogue. We find that over the range 1$\le l\le$100 
the 2MASS $C_l$ is steeper than that for the Hubble Volume model. However, in the linear regime ($l\le$30) there is good agreement between the two. We 
investigate in detail the effects of possible sources of systematic error. Converting linear power spectrum predictions for the form of the 
three-dimensional matter power spectrum, $P(k)$, and assuming a flat CDM cosmology, a primordial $n_s$=1 spectrum and negligible neutrino mass, we 
perform fits to the galaxy angular power spectrum at large linear scales ($l\le$30, corresponding to $r$\gsim 50\mpc). We obtain 
constraints on the galaxy power spectrum shape of $\Gamma_{eff}=0.14\pm0.02$, in good agreement with previous estimates inferred at smaller scales. 
We also constrain the galaxy power spectrum normalisation to ($\sigma_8b_K$)$^2=1.36\pm0.10$; in combination with previous constraints on $\sigma_8$ 
we infer a $K_s$-band bias of $b_K$=1.39$\pm$0.12. We are also able to provide weak constraints on $\Omega_m h$ and $\Omega_b$/$\Omega_m$. These 
results are based on the usual assumption that the errors derived from the Hubble Volume mocks are applicable to all other models. If we instead 
assume that the error is proportional to the $C_l$ amplitude then the constraints weaken; for example it becomes more difficult to reject cosmologies 
with lower $\Gamma_{eff}$.
\end{abstract}


\section{Introduction}

The nature of galaxy fluctuations at extremely large scales ($r$\lsim 1000 \mpc ) is poorly constrained. Over the last decade, large galaxy 
surveys have constrained the form of the galaxy density field to a few hundred Megaparsecs. However, the agreement with the concordance model at 
these scales can only be weakly inferred. Indeed, recent evidence has suggested that there may be excess power over the expected $\Lambda$CDM form to 
the three-dimensional power spectrum of matter, $P(k)$, at large scales \citep{fri4,fri3,fri,bus} arising from large inhomogeneities in the local 
galaxy distribution. 

In recent years, large redshift surveys of both galaxies \citep{cole2,per,zeh} and QSOs \citep{out} have determined 
$P(k)$ at relatively small scales. Using the 2dF Galaxy Redshift Survey (2dFGRS), \citet{cole2} have constrained the form 
of galaxy density fluctuations to scales of $r\approx$300\mpc ~and the associated cosmological parameters to 
$\Omega_m h$=0.168$\pm$0.016 and $\Omega_b$/$\Omega_m$=0.185$\pm$0.046 (assuming $h$=0.72). However, determining the power 
spectrum through such redshift surveys suffers from large statistical uncertainty at large scales due to the relatively few objects available, as well 
as uncertainties arising from cosmic variance due to the relatively small volumes surveyed. 

\begin{figure*}
\begin{center}
\centerline{\epsfxsize = 7in
\epsfbox{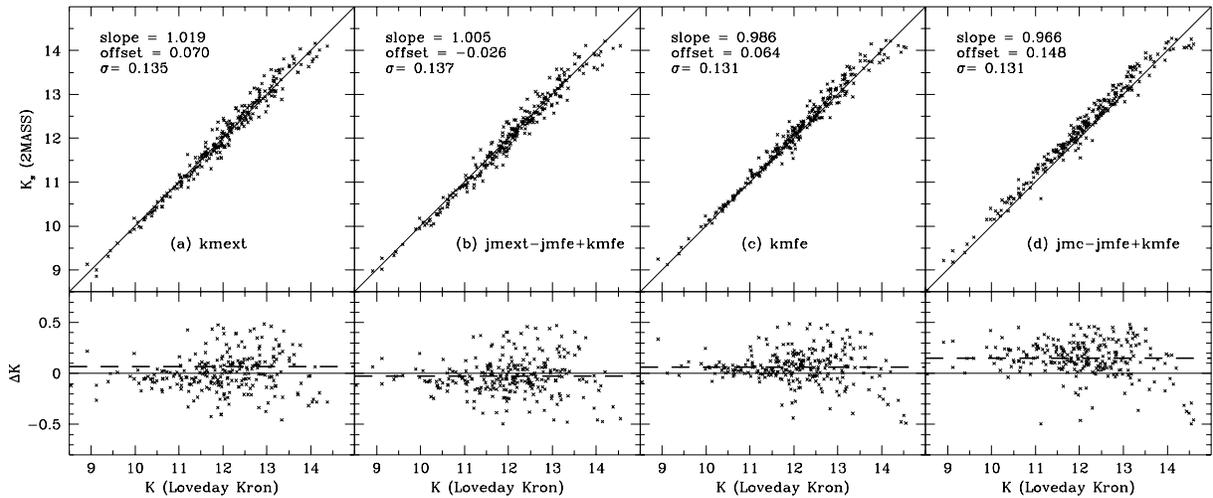}}
\caption{A selection of $K_s$-band magnitude estimates from the 2MASS full release compared with \citet{lov} $K$-band
photometry. In each case the lower panels display the residual. The $x=y$ slope is indicated by a solid line, while
the mean offset is indicated in the lower panel by a dashed line. This offset (in magnitudes), the best fit slope
determined from least squares fits and the $rms$ scatter are indicated for each magnitude estimate. We estimate
the magnitudes directly from the (a) $K_s$-band extrapolated and (c) $K_s$-band fiducial elliptical Kron magnitudes, and also
from the (b) $J$-band extrapolated and (d) $J$-band circular Kron magnitudes colour-corrected to the $K_s$-band using
the $J$ and $K_s$-band fiducial elliptical Kron magnitudes.}
\label{fig:2massphot}
\end{center}  
\end{figure*}

Using imaging surveys as opposed to redshift surveys provides a greater number of objects over larger solid angles. With angular power spectrum 
analysis of such surveys it is therefore possible to constrain the form of galaxy fluctuations to extremely large scales. However, since the 
clustering signal from a particular scale in real space is smeared over a range of angular scales, cosmological constraints through comparisons with 
linear theory predictions at smaller scales cannot be made; the departure from linearity at scales of $r$\lsim 40\mpc ~\citep{per} affects 
the clustering signal in the angular power spectrum over a wide range of scales. Nevertheless at large scales, where this effect is 
insignificant, angular power spectrum analysis represents one of the most effective probes of local large-scale structure.

Previously, the galaxy angular power spectrum has been determined for the Sloan Digital Sky Survey Early Data Release, the Edinburgh-Durham 
Southern Galaxy Catalogue, and a sample of IRAS galaxies \citep[~respectively]{teg,hut,sch}, which along with the recent 
analyses of redshift surveys has constrained the form of galaxy fluctuations to $r\approx300$\mpc\ .

The 2 Micron All Sky Survey (2MASS) has recently been completed and provides $K_s$, $H$ and $J$-band photometry for $\approx$1.6$\times$10$^6$ 
extended sources over the entire sky to $K_s$\gsim13.5 \citep{jar2,jar}; at the time of writing this dataset represents the largest all sky galaxy 
survey. The 2MASS data therefore represents a uniquely powerful probe of the local 
galaxy density field at large scales; applying a galactic latitude cut of $|b|>$20$^{\circ}$ in order to remove regions of high extinction 
and stellar contamination yields a sample containing 518~576 $K_s<13.5$ galaxies, probing a volume approximately 5 times larger than 
the final 2dF Galaxy Redshift Survey (2dFGRS) volume. A further advantage of 2MASS over previous datasets is that the photometry is extremely 
accurate with high completeness; the photometric zero-point calibration is accurate to 2-3\%; galaxy identification is $\approx$99\% reliable and the 
galaxy catalogue is $>$90\% complete for $|b|>$20$^{\circ}$ \citep{jar2}. 
 
In this paper, we use data from the 2MASS final release extended source catalogue to determine the $K_s$-band galaxy 
angular power spectrum with the aim of determining the form of the clustering of galaxies at extremely large scales, and 
constraining the shape and normalisation of the power spectrum. In section 2, we describe the 2MASS dataset and the magnitude 
estimator used. The method of analysis is outlined and the 2MASS angular power spectrum is determined and 
compared to mock power spectra in section 3. In section 4, we investigate various sources of 
systematic error. We determine constraints for various cosmological parameters in section 5. The conclusions follow in section 6.

\section{Data}

\subsection{The 2MASS Extended Source Catalogue}

The 2 Micron All Sky Survey (2MASS) final release extended source catalogue provides $K_s$, $H$ and $J$-band photometry for over 
1.6$\times$10$^6$ extended sources over the entire sky with high completeness to $K_s$=13.5 \citep{jar2}. 

Previously, in order to estimate the total $K_s$-band magnitudes from the 2MASS second incremental release data, \citet{cole} used
the deeper $J$-band Kron magnitudes, colour-corrected to the $K_s$-band via the $J$ and $K_s$ default aperture magnitudes. The accuracy of this 
magnitude estimator was determined through a comparison with the $K$-band photometry of \citet{lov}; the Loveday photometry had 
better signal-to-noise and resolution than the 2MASS scans and so enabled more accurate 2MASS magnitudes to be determined.

The final release data uses revised magnitude estimates and the default aperture magnitudes used in \citet{cole} have been abandoned 
(Jarrett - priv. comm.). In Fig.~\ref{fig:2massphot} we show a selection of 2MASS $K_s$-band magnitude estimates with the revised 2MASS 
photometry compared with the \citet{lov} photometry used previously. 
In the place of the default aperture magnitudes used in \citet{cole}, we use fiducial elliptical Kron magnitudes in panels (b) and (d) 
to colour-correct the $J$-band magnitudes to the $K_s$-band. Of the many different magnitude estimates examined, the most accurate 
in terms of the scale error between the Loveday and 2MASS photometry and the zero-point offset uses the $J$-band 
extrapolated magnitude colour-corrected to the $K_s$-band as described above. Using the dust maps of \citet{shl}, the main galaxy sample uses 
extinction-corrected $K_s$-band magnitudes calculated in this way. 

In order to verify the usefulness of the magnitude estimator used in this work as an estimate of the total $K_s$-band magnitude, 
we perform an internal check via a comparison with the magnitude estimates used in the 2MASS-selected 6dF Galaxy Survey (6dFGS). The 6dFGS 
$K_s$-band magnitudes are determined using a surface brightness correction to the $K_s$-band 20 mag. arcsec$^{-2}$ isophotal 
elliptical aperture magnitude \citep{hea}. We find excellent agreement with a slope of 1.022, an offset of 0.018 magnitudes and a spread of 
$\sigma$=0.048 magnitudes for $|b|>$20$^{\circ}$ galaxies matched below $K_s=13.5$.

The 2MASS dataset removes or flags sources identified as artefacts such as diffraction spikes and meteor streaks \citep{jar}; we use the 2MASS 
$cc\_flag$ to remove such objects. We also employ a colour cut ($J-K_s<$0.7 and $J-K_s>$1.4) below $K_s$=12 in order to remove a small number 
of objects identified as non-extragalactic extended sources \citep{mal2,mal}. In this work, our main sample includes 518~576 
$K_s<$13.5 galaxies above a galactic latitude of $|b|=$20$^{\circ}$. For reference, the surface density 
is 19.1 deg$^{-2}$. We also use a shallower sample limited at $K_s$=12.5 and  $|b|>$20$^{\circ}$ which includes 124~264 galaxies and for which the 
surface density is 4.58 deg$^{-2}$. 

\subsection{The $\Lambda$CDM Hubble Volume Simulation}

The Hubble Volume catalogues represent the largest volume N-body simulations of the Universe to date. The $\Lambda$CDM simulation
follows the evolution of 10$^9$ dark matter particles from $z\approx$50 over a volume of 3000$^3~h^{-3}$Mpc$^3$ to a resolution of 
$\approx$3\mpc. The associated cosmological parameters are $\Omega_m$=0.3, $\Omega_b$=0.04, $h$=0.7, $\sigma_8$=0.9 \citep{jen}.

In this work, we construct mock 2MASS catalogues from the $z=0$ $\Lambda$CDM Hubble Volume simulation dark matter particles. We divide the 
total volume into 27 virtually independent spherical volumes of $r=500$\mpc . These are subjected to the 2MASS selection function:

\begin{equation}
n(z)=\frac{3z^2}{2(\bar{z}/1.412)^3} exp \left(-\left(\frac{1.412z}{\bar{z}}\right)^{3/2}\right)
\label{equation:sel}
\end{equation}

\noindent \citep{bau,mal} where $\bar{z}$ is determined from the 2MASS-2dFGRS matched sample described in \citet{fri4}; for reference 
$\bar{z}$=0.074 for $K_s<$13.5 and $\bar{z}$=0.050 for $K_s<$12.5. Equation~\ref{equation:sel} is normalised to match the total number of observed 
2MASS galaxies for $|b|>$20$^{\circ}$. Due to the volume of the 27 mock 2MASS catalogues, the selection function is artificially truncated for
the $K_s<$13.5 mocks at $z\approx$0.156. However, this has a negligible effect on the work in this paper; at this redshift,
$\approx$95\% of the galaxies are sampled for $K_s<$13.5.

\section{The 2MASS Angular Power Spectrum}

\subsection{Estimating the Power Spectrum}

\begin{figure*}
\begin{center}
\centerline{\epsfxsize = 7in
\epsfbox{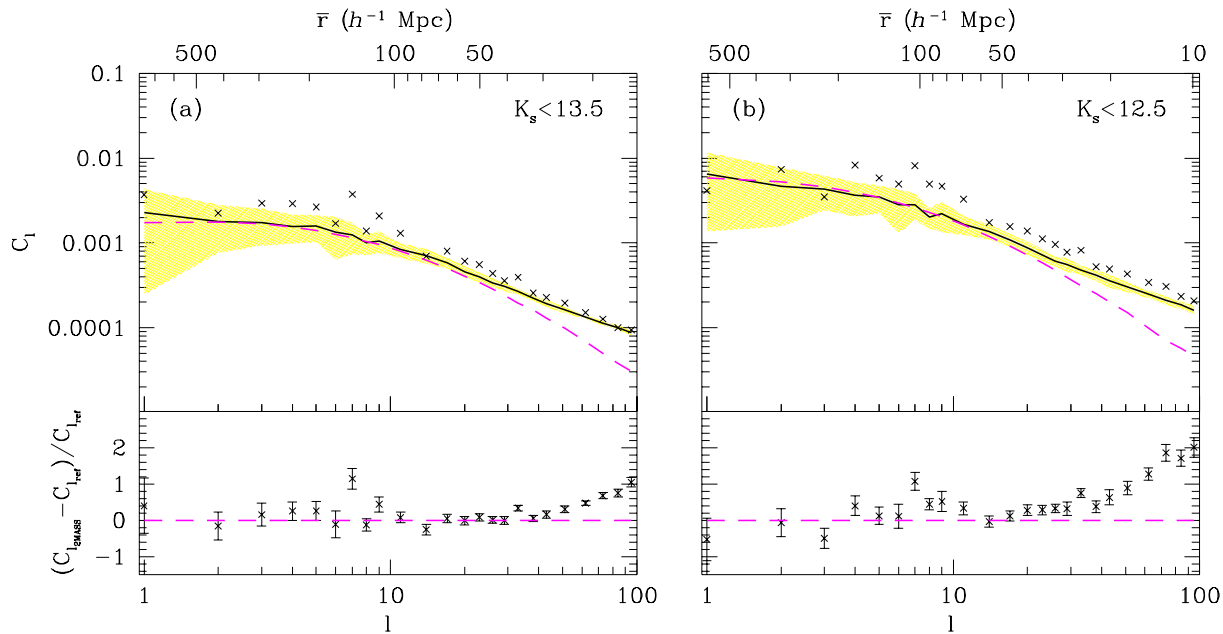}}
\caption{The $|b|>$20$^{\circ}$ 2MASS galaxy angular power spectrum for (a) 518~576 $K_s<13.5$ and (b) 124~264 $K_s<12.5$ galaxies.
The crosses indicate the 2MASS datapoints with the shaded region and solid line indicating the 1$\sigma$ spread and mean power spectrum of
the 27 mock unbiased 2MASS catalogues constructed from the $\Lambda$CDM Hubble Volume mock catalogue as described in section 2.2. In each
case, unbiased linear theory models corresponding to the Hubble Volume mock catalogue input parameters of $\Omega_m$=0.3, $\Omega_b$=0.04, $h$=0.7 
and $\sigma_8$=0.9 are indicated by the dashed lines. In the lower panels we show the fractional deviation of the 2MASS power spectrum from
this model applying the best fit power spectrum normalisation, $\sigma_8b_k^2$=1.36, (determined in section 5 for the $K_s<$13.5 sample) to the 
linear prediction, with errors taken from the mock 2MASS 1$\sigma$ spread. In addition we indicate the approximate mean distance scale probed by the 
data for each $l$-mode on the top $x$-axis.}
\label{fig:2mass_mag}
\end{center}
\end{figure*}

\begin{figure*}
\begin{center}
\centerline{\epsfxsize = 7in
\epsfbox{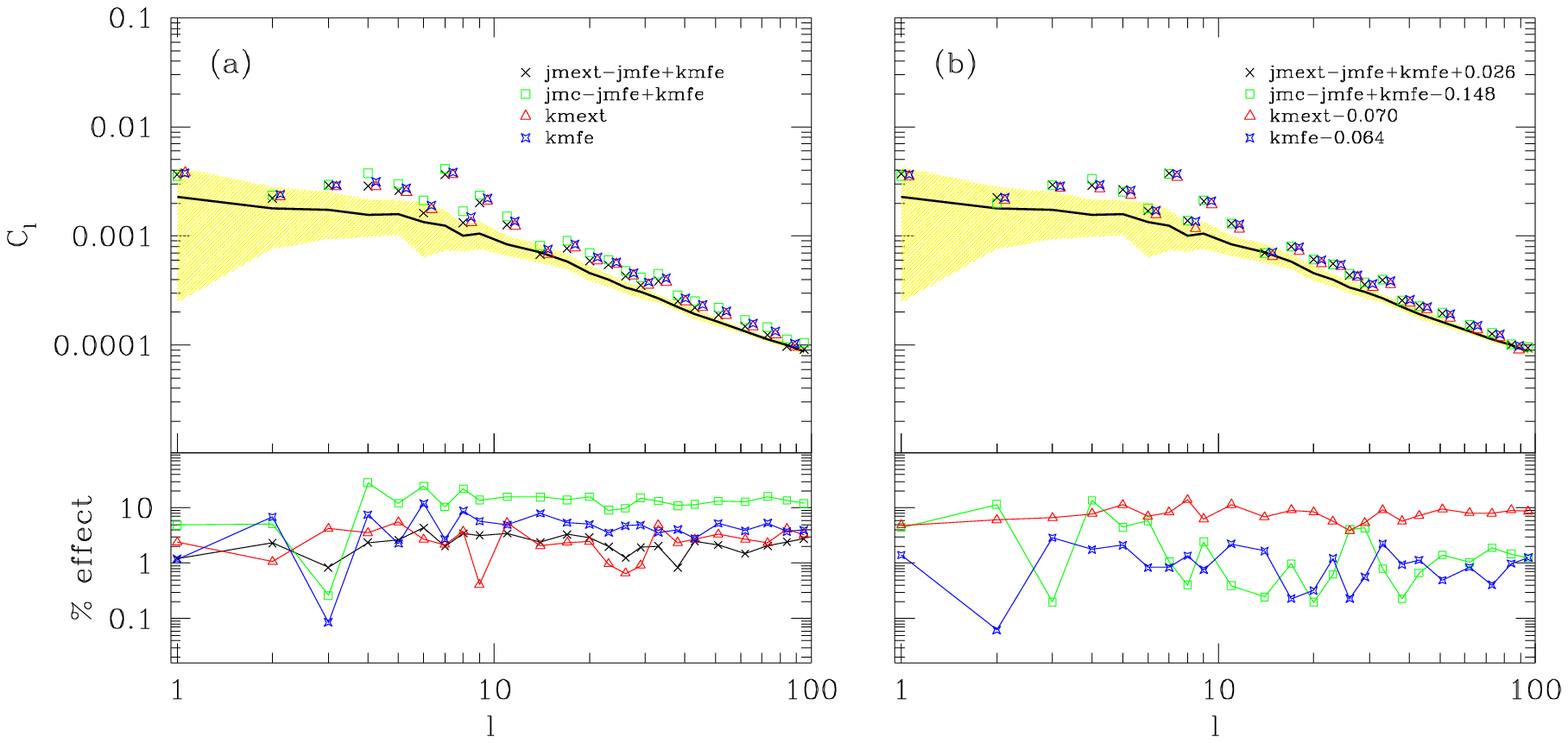}}
\caption{The $|b|>$20$^{\circ}$, $K_s<$13.5 extinction-corrected 2MASS galaxy angular power spectra for the four magnitude estimators
shown in Fig.~\ref{fig:2massphot} using (a) the raw magnitude estimate and (b) a zero-point correction to account for the offset
determined with respect to the \citet{lov} photometry. The 1$\sigma$ spread and mean 2MASS mock power spectrum are shown as in
Fig.~\ref{fig:2mass_mag}. The lower panels indicate the effect of each magnitude estimator on the resulting power spectrum compared to the
colour-corrected $J$-band extrapolated magnitude estimator (with the zero-point correction) used in Fig.~\ref{fig:2mass_mag} and also
indicated here by the black crosses in panel (b). In the upper panels we have displaced the kmext and kmfe datapoints for clarity.}
\label{fig:2mass_mag2}
\end{center}
\end{figure*}

Following the usual method \citep[e.g.][]{peb,peb2,peb3,sch}, the angular power is estimated 
through a spherical harmonic expansion of the surface density of galaxies. The coefficients of this expansion are determined 
over the observed solid angle $\Omega_{obs}$: 

\begin{equation}
a_l^m =\sum_{N_{gal}} Y_l^m(\theta ,\phi ) - {\mathcal{N}} \int_{\Omega_{obs}} Y_l^m(\theta ,\phi ) d\Omega
\end{equation}

\noindent where ${\mathcal{N}}$=$N_{gal}/\Omega_{obs}$ is the observed number of galaxies per steradian. The angular power
is then determined:

\begin{equation}
{\mathcal{Z}}_l=\frac{1}{2l+1} \sum_m \frac{|a_l^m|^2}{{\mathcal{J}}_l^m}
\end{equation}

\noindent where,

\begin{equation}
{\mathcal{J}}_l^m = \int_{\Omega_{obs}} Y_l^m(\theta ,\phi ) d\Omega
\end{equation}

\noindent The angular power is then normalised, subtracting the expected shot noise contribution:

\begin{equation}
C_l = \frac{{\mathcal{Z}}_l}{{\mathcal{N}}} - 1
\label{equation:norm}
\end{equation}

\noindent such that $C_l$=0 corresponds to a random distribution. 

\subsection{Fitting to the Power Spectrum}
In order to compare the angular power spectrum with cosmological predictions, we determine an expected form for the 
angular power spectrum for various cosmological parameters using the relation between the three and two-dimensional power spectra:

\begin{equation}
|a_l^m|^2 = \frac{2}{\pi}\int \left( \int r^2 \Phi(r) j_l (kr) dr \right)^2 k^2 P(k) dk + {\mathcal{N}}
\label{equation:2d}
\end{equation}

\noindent \citep{sch,teg,hut}, which we normalise as before. Here, $\Phi(r)$ is the 2MASS selection 
function, and $j_l$ is a spherical Bessel function. The 2MASS selection function is determined using equation 1.

We use the transfer function fitting formulae of \citet{eis} to obtain a linear theory prediction for the dark matter power spectrum, $P(k)$, with 
input parameters for the matter, vacuum, baryon and neutrino densities ($\Omega_m$, $\Omega_{\Lambda}$, $\Omega_b$ and $\Omega_{\nu}$), $h$ (such that 
$H_0=100h$) and matter power spectrum normalisation ($\sigma_8$). We also employ a linear biasing scheme such that 
$P_{gal}(k)$=$b^2P_{matter}(k)$ to provide a linear prediction for the galaxy $P(k)$. This is then transformed to a galaxy angular power spectrum 
prediction using the spherical Bessel function transform in equation~\ref{equation:2d}.

\subsection{Results}

\begin{figure}
\begin{center}
\centerline{\epsfxsize = 3.5in
\epsfbox{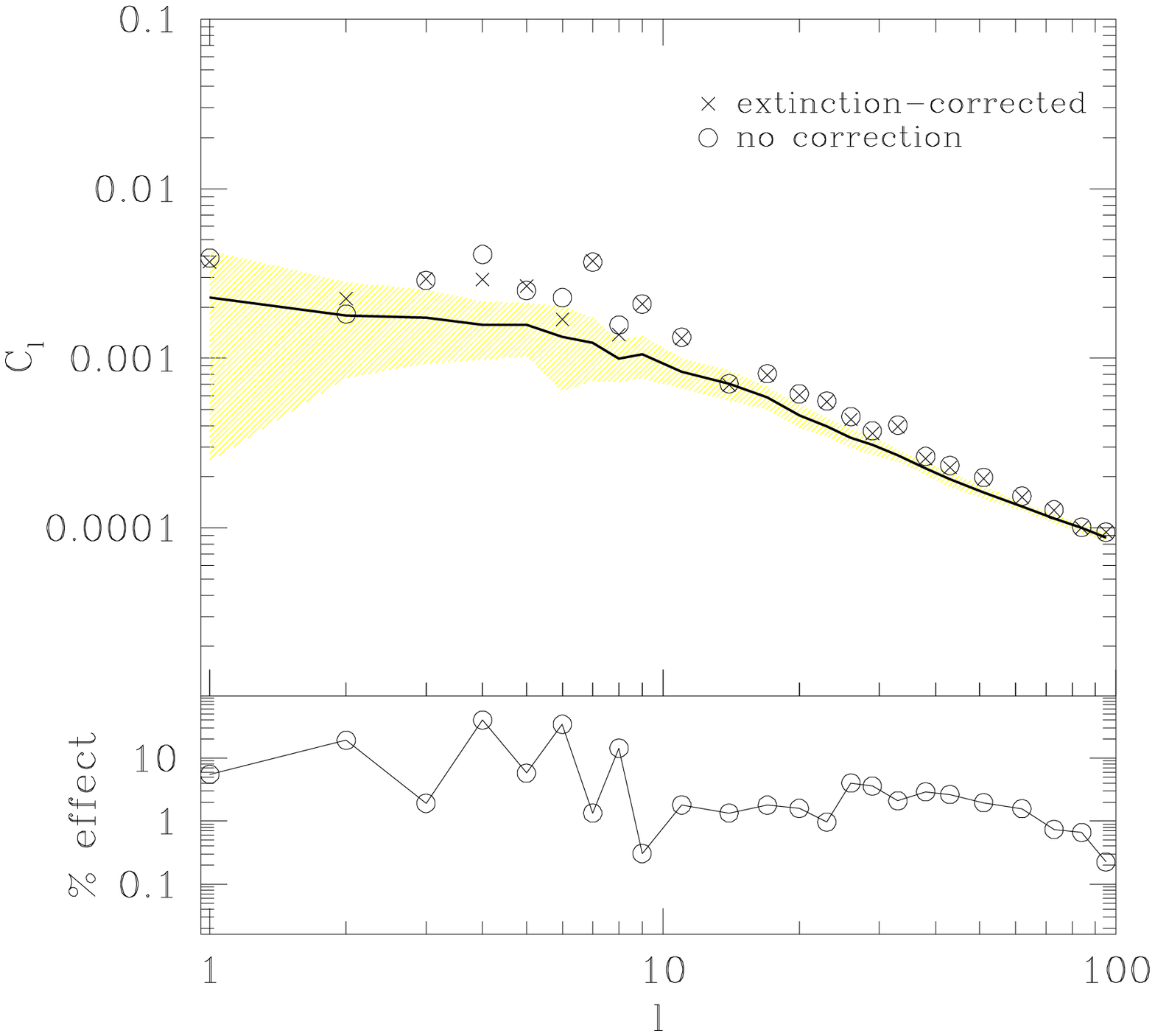}}
\caption{The $|b|>$20$^{\circ}$, $K_s<$13.5 2MASS galaxy angular power spectra including no extinction correction, and as previously 
an extinction correction derived from the \citet{shl} dust maps. The mock 2MASS mean angular power spectrum and 1$\sigma$ spread are shown 
as before. In the lower panel we indicate the effect of this correction on the power spectrum through a comparison with the corrected 
sample (indicated by the crosses in the upper panel and as shown in Fig~\ref{fig:2mass_mag}a).}
\label{fig:2mass_ext}
\end{center}
\end{figure}

\begin{figure}
\begin{center}
\centerline{\epsfxsize = 3.5in
\epsfbox{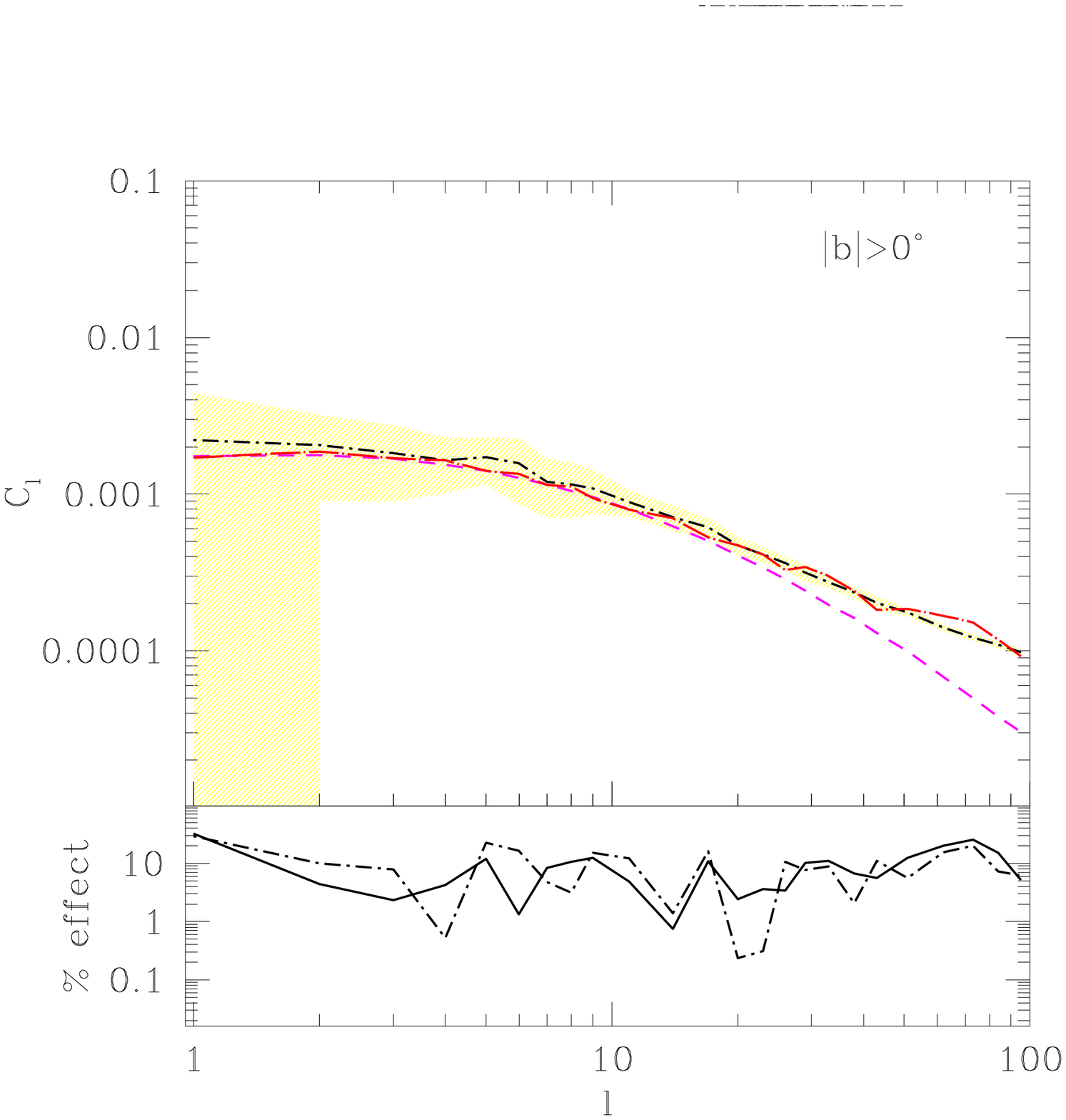}}
\caption{The $|b|>$0$^{\circ}$, $K_s<$13.5 mean power spectrum and 1$\sigma$ spread determined from the 27 mock 2MASS catalogues (solid line and 
shaded region). As in Fig.~\ref{fig:2mass_mag}a the dashed line indicates the expected linear trend for the $\Lambda$CDM Hubble Volume mock input
parameters of $\Omega_m$=0.3, $\Omega_b$=0.04, $h$=0.7 and $\sigma_8$=0.9. As a consistency check, we also show the $\Lambda$CDM Hubble
Volume non-linear power spectrum (large dot-dashed line) calculated via the numerically-determined $\Lambda$CDM Hubble Volume $P(k)$
(Carlton Baugh - priv. comm.) transformed to the angular power spectrum as described in section 3.2. In the lower panel we compare this
prediction with the $|b|>$0$^{\circ}$ (dot-dashed line) and $|b|>$20$^{\circ}$ (solid line) mean mock 2MASS power spectra.}
\label{fig:2mass_b0}
\end{center}
\end{figure}

The angular power spectrum for 518~576 $K_s<13.5$, $|b|>$20$^{\circ}$ 2MASS galaxies is presented in 
Fig.~\ref{fig:2mass_mag}a, determined through a spherical harmonic expansion of the galaxy number density as described in section 3.1. In order to 
determine the expected scatter due to cosmic variance we determine the angular power spectrum for the 27 unbiased mock 2MASS catalogues constructed 
from the $\Lambda$CDM Hubble Volume simulation described in section 2.2; the mean and 1$\sigma$ spread are indicated by the solid line and shaded 
region. On the top $x$-axis we also indicate the approximate distance scale probed by the angular power spectrum at the mean depth of the sample 
determined from the 27 mock 2MASS catalogues. At the very smallest $l$-modes, the $K_s<$13.5 power spectrum probes scales of \gsim500\mpc.

We have also calculated the linear prediction 
corresponding to the $\Lambda$CDM Hubble Volume input parameters ($\Omega_m$=0.3, $\Omega_{\Lambda}$=0.7, $\Omega_b$=0.04, $h$=0.7, 
$\sigma_8$=0.9 and $\Omega_{\nu}$=0) through a spherical Bessel function transform of the three-dimensional 
power spectrum as described in section 3.2; this is indicated for a bias of 1.0 by the dashed line. The linear model and the 
mean mock 2MASS power spectrum are in good agreement at large scales. At smaller angular scales ($l>$30) the effects of non-linear clustering 
become significant.

In order to verify whether the form and scatter of the mock power spectra, which we later use to estimate the error on the observed angular power 
spectrum, is consistent with the data, we perform a $\chi^2$ fit between the two. We marginalise over the normalisation of the mean mock angular power 
spectrum and use the binning as shown in order to reduce the covariance to insignificant levels. We assume that the spread in the mock power 
spectra is independent of normalisation, i.e. we apply the same spread determined for the unbiased mock power spectra to 
the observed angular power spectrum. In this particular case, this is likely to provide an optimistic view of the observed errors since we 
are not shot noise limited. In this scenario, the errors are likely to be independent of the power spectrum amplitude; on the other hand, 
if the observed power spectrum is cosmic variance limited the errors scale with model normalisation (see \citet{fel} for further discussion 
on this point). We investigate the impact of this assumption on the associated cosmological constraints in section 5. First, we perform a $\chi^2$ 
fit over the full angular range 1$\le l\le$100 between the $K_s<$13.5 2MASS galaxy angular power spectrum and the mean mock 2MASS power spectrum; we 
find that $\chi^2/d.o.f.$=3.0. Limiting the angular range to scales which are not significantly affected by 
non-linear clustering ($l\ge$30), the form of the mock power spectra are in better agreement with the observed 2MASS galaxy angular power spectrum, 
with $\chi^2/d.o.f.$=2.0.

The form of the 2MASS angular power spectrum is therefore in good agreement with the $\Lambda$CDM prediction in the linear regime, although it is 
clear from Fig.~\ref{fig:2mass_mag}a that there is some difference in slope at small scales. Assuming the validity of the prediction, this is due 
either to scale-dependent bias in the non-linear regime or resolution effects in the Hubble Volume simulation. Consistency with the $\Lambda$CDM
prediction in the linear regime, of interest in this work, is confirmed through a comparison (in the lower panel) with the linear prediction for the 
$\Lambda$CDM Hubble Volume simulation input parameters applying a scale-independent bias to match the normalisation of the observed power spectrum at
large scales (see section 5).

\section{Systematic Errors}

\begin{figure}
\begin{center}
\centerline{\epsfxsize = 3.5in
\epsfbox{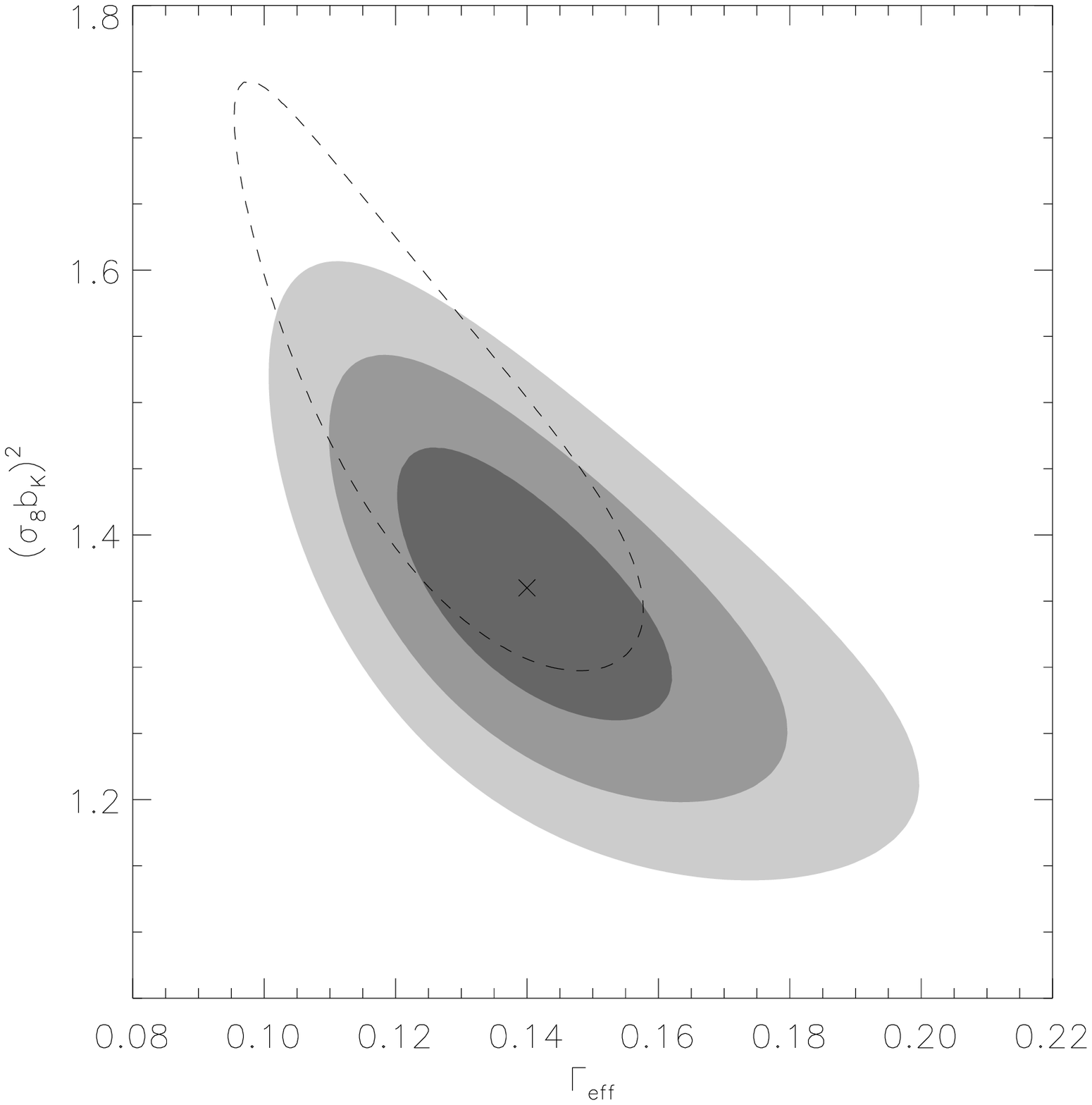}}
\caption{Filled contours representing the 1$\sigma$, 2$\sigma$ and 3$\sigma$ confidence regions for the galaxy power spectrum
shape and normalisation determined from $\chi^2$ fits to the 2MASS $|b|>$20$^{\circ}$ $K_s<$13.5 galaxy angular power spectrum in the
range $l\le$30. The cross indicates the best fit parameters of $\Gamma_{eff}$=0.14 and ($\sigma_8b_K$)$^2$=1.36. We also show the 1$\sigma$ confidence 
region for the 2MASS result as above where we use errors which scale with the model power spectrum normalisation (dashed line).}
\label{fig:sigbb}
\end{center}
\end{figure}

\begin{figure}
\begin{center}
\centerline{\epsfxsize = 3.5in
\epsfbox{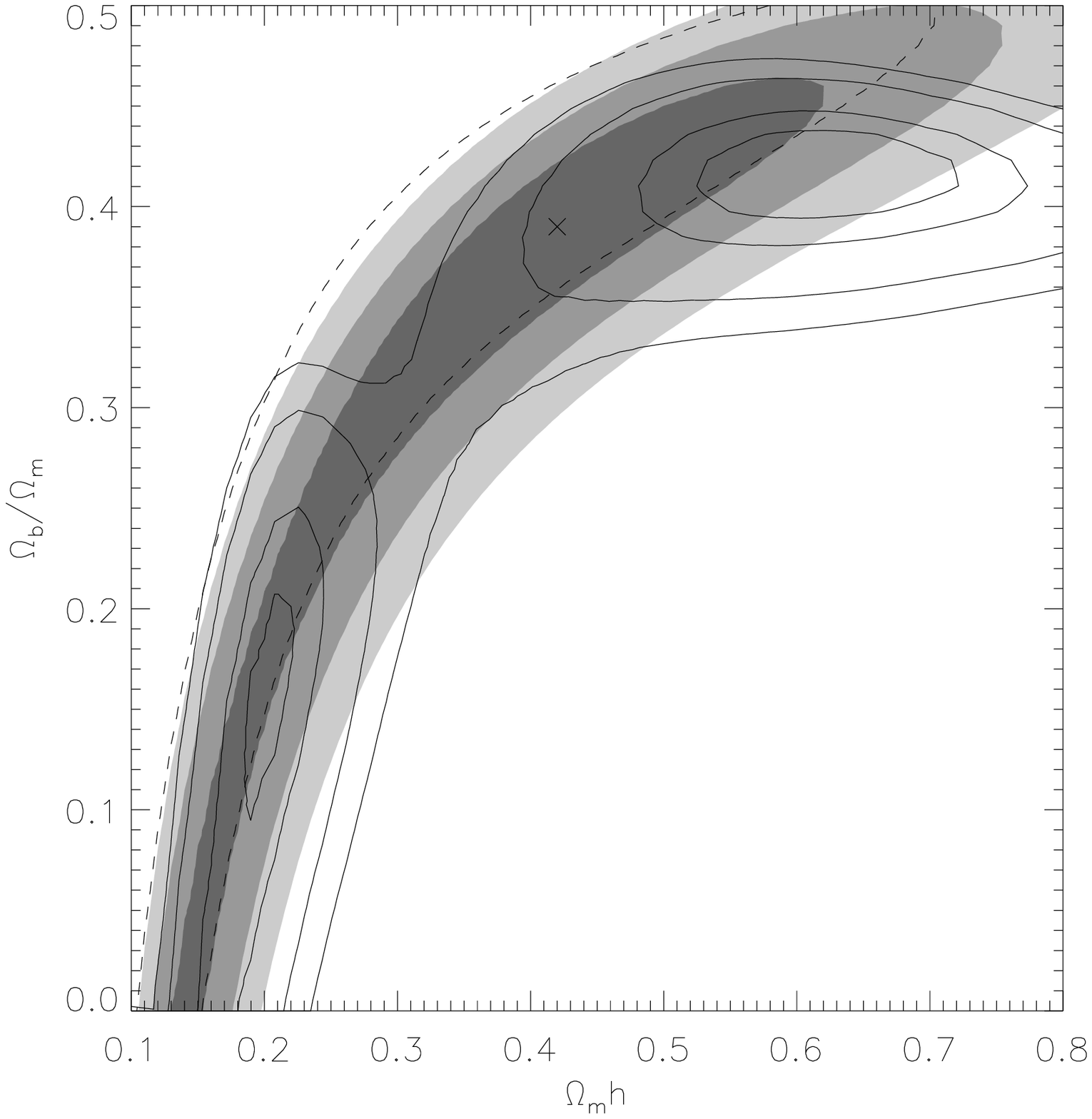}}
\caption{Contours of decreasing likelihood in the $\Omega_m h$ - $\Omega_b$/$\Omega_m$ plane for the best-fitting
angular power spectrum in the range $l\le$30. The filled contours indicate the 1$\sigma$, 2$\sigma$ and 3$\sigma$
confidence regions for the 2MASS $|b|>$20$^{\circ}$ $K_s<$13.5 galaxy angular power spectrum, determined from simple $\chi^2$ fits,
marginalising over the normalisation and $h$. The solid contours indicate the 1$\sigma$, 2$\sigma$, 3$\sigma$ and
4$\sigma$ confidence regions determined from the 2dFGRS 100k release $P(k)$ \citep{per}. The cross marks the best fit model to the 2MASS
data of $\Omega_m h$=0.42 and $\Omega_b$/$\Omega_m$=0.39. As in Fig.~\ref{fig:sigbb} we also show the 1$\sigma$ confidence region for the 2MASS result 
as above where we use errors which scale with the model power spectrum normalisation (dashed line).}
\label{fig:conf}
\end{center}
\end{figure}

\subsection{Magnitude Limits}

Before turning to the cosmological constraints inferred from the 2MASS galaxy angular power spectrum it is important to verify that the 
results are robust and not significantly affected by potential sources of systematic error. While the 2MASS catalogue is $>$98\% reliable 
for $|b|>$20$^{\circ}$, $K_s<$13.5 galaxies \citep{jar} and 99\% complete for $|b|>$30$^{\circ}$, 12.0$<K_s<$13.7 galaxies 
\citep{mal}, we wish to verify that the angular power spectrum is robust to changes in the magnitude limit, and is not 
adversely affected by variable incompleteness or reliability at faint magnitudes or scale errors in the photometry.

Fig.~\ref{fig:2mass_mag} shows the 2MASS galaxy angular power spectrum as a function of imposed magnitude limit. 
The shape and normalisation of the power spectrum, with respect to both the linear model and the mean mock 2MASS power spectrum, are 
remarkably robust to changes in the magnitude limit. The departure of the linear model from the observed power spectrum occurs at larger 
angular scales with the shallower magnitude limit due to the reduced mean depth of the sample. For this reason also, the mock 2MASS power 
spectrum is more significantly distorted at the very smallest scales by resolution affects resulting in a slightly steeper slope at 
$l$\gsim70. 

\subsection{Magnitude Estimator} 
Throughout this paper, we estimate the $K_s$-band magnitudes using the $J$-band extrapolated magnitudes colour-corrected using 
the $K_s$ and $J$-band fiducial elliptical Kron magnitudes, as this results in a smaller zero-point offset and scale error 
when compared to the more accurate $K$-band photometry of \citet{lov}. We wish to investigate the effect on the power 
spectrum by the choice of magnitude estimator; in Fig.~\ref{fig:2mass_mag2}a and b we compare the power spectra for the four 
magnitude estimators presented in Fig.~\ref{fig:2massphot} with and without respectively the associated 
correction to the \citet{lov} zero-point. 

The power spectrum is robust to changes in the magnitude estimate and zero-point at the \lsim10\% level. This is due to 
the fact that the change in the depth of the survey due to differences in the magnitude limit and scale error effects are 
insignificant. 

\subsection{Extinction}

While the level of extinction in the $K_s$-band is low and the 2MASS magnitudes have been corrected using the \citet{shl} 
dust maps, it is useful to examine the potential level of systematic error introduced by extinction. Fig.~\ref{fig:2mass_ext} 
shows the 2MASS galaxy angular power spectrum with and without correction for extinction. In this extreme case, the effect of removing the dust 
correction to the magnitude estimate is at the \lsim10\% level at large scales and \lsim1\% above $l\approx$10.

\subsection{The Window Function}
Throughout this paper a $|b|>$20$^{\circ}$ galactic latitude cut is applied in order to avoid the high levels of
extinction and stellar contamination in the zone of avoidance. We wish to determine the level of any systematic effect on the spread 
of the Hubble Volume mock power spectra (and therefore our interpretation of the statistical uncertainty) introduced by the window 
function. In Fig.~\ref{fig:2mass_b0} the 27 mock 2MASS power spectra and corresponding linear theory model for the $\Lambda$CDM Hubble Volume 
input parameters are shown with no galactic latitude cut. Neither the shape nor the spread of the power spectra are significantly altered. The effect 
of the window function on the angular power spectrum is \lsim5\% at all scales. 

In order to check the consistency of our results we provide a further verification of the mock 2MASS power spectrum results through a 
comparison with the transform of the numerically-determined $\Lambda$CDM Hubble Volume simulation $P(k)$ (Carlton Baugh - priv. comm.). There is 
good agreement with both the $|b|>$0$^{\circ}$ and $|b|>$20$^{\circ}$ mean mock 2MASS power spectra.

\section{Cosmological Constraints}

Using the 2MASS galaxy angular power spectrum we have determined the form of the galaxy density 
field at extremely large scales and verified that it is not significantly affected by common sources of systematic error. 
We now wish to determine the associated cosmological constraints. 

Using the \citet{eis} transfer function fitting formulae we have determined linear theory predictions for the three-dimensional power spectrum of 
matter,  $P(k)$, using input parameters of $\Omega_m$, $\Omega_{\Lambda}$, $\Omega_b$, $h$ and matter power spectrum normalisation, $\sigma_8$; 
in the subsequent analysis we assume a negligible neutrino mass density, a primordial $n_s$=1 spectrum and $\Omega_{\Lambda}$=1-$\Omega_m$. We form 
galaxy angular power spectrum predictions using the spherical Bessel function transform described in section 3.2 and a linear biasing scheme. 

First, we perform fits to the galaxy power spectrum shape and normalisation. Assuming a CDM 
cosmology, the power spectrum can be defined through a parameterisation of the shape

\begin{equation}
\Gamma_{eff}=\Omega_mh\: exp(-\Omega_b(1+\sqrt[]{2h}/\Omega_m))
\end{equation}

\noindent \citep{sug}, and a normalisation, which for galaxy power spectra may be parameterised through the galaxy bias and $\sigma_8$. 
We constrain $\Gamma_{eff}$ and ($\sigma_8b_K$)$^2$ using a grid of 200$\times$800 models between 0.1$\le\Gamma_{eff}\le$0.3 and 
0.0$\le$($\sigma_8b_K$)$^2\le$8.0 respectively. We perform least squares fits to the 
$|b|>$20$^{\circ}$, $K_s<$13.5 angular power spectrum as shown in Fig.~\ref{fig:2mass_mag}a at scales of $l\le$30 (binned as shown to reduce the 
covariance to insignificant levels); beyond $l\approx$30 the angular power spectrum begins to be significantly affected by non-linear effects. 

We take the spread determined from the 27 mock 2MASS angular power spectra in order to estimate the errors on the 2MASS datapoints, assuming that 
the uncertainty remains the same for a biased as for an unbiased distribution (as in section 3.3). In doing this, we assume that 
the $\Lambda$CDM Hubble Volume mock catalogue provides an accurate description of the local galaxy distribution at large scales and that the 
associated uncertainty in the datapoints is realistic. However, since these errors are valid only in an unbiased $\Lambda$CDM cosmology we are 
required to make assumptions as to the nature of the cosmic variance in the various other cosmologies scrutinised in these fits. Here we assume that 
the errors are independent of cosmology and power spectrum normalisation; the likely impact of this assumption is examined below. We find that: \\
\\
\noindent $\Gamma_{eff}=0.14\pm0.02$
\\
\\
\noindent and
\\
\\
\noindent ($\sigma_8b_K$)$^2=1.36\pm0.10$
\\
\\
marginalising over the normalisation and power spectrum shape respectively. The associated confidence regions are indicated by the filled contours in 
Fig.~\ref{fig:sigbb}. 

This value of $\Gamma_{eff}$ is in excellent agreement with the 2dFGRS fit \citep{per} of $\Gamma_{eff}=0.18\pm0.04$ (for $h$=0.7) and 
the WMAP value \citep{spe} of $\Gamma_{eff}=0.15\pm0.01$ (for $n_s$=0.99). However, our value is 
slightly higher than the \citet{mal} result which constrains $\Gamma_{eff}=0.116\pm0.009$ at 95\% confidence using a measure 
of the three-dimensional $K_s$-band galaxy power spectrum via an inversion of the 2MASS angular correlation function. 

\begin{figure}
\begin{center}
\centerline{\epsfxsize = 3.5in
\epsfbox{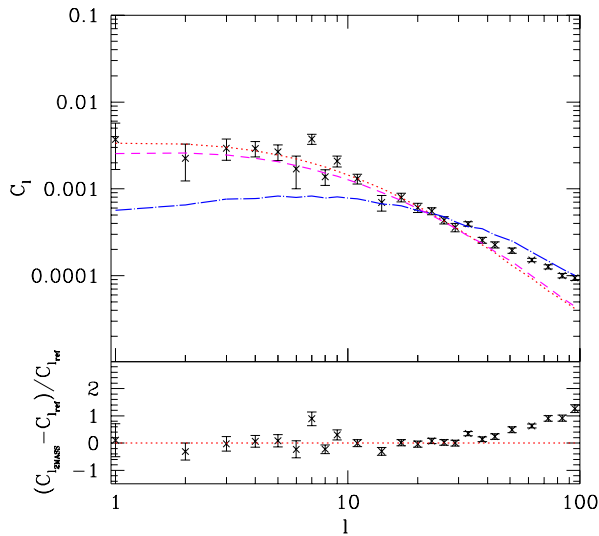}}
\caption{The angular power spectrum for $|b|>$20$^{\circ}$ $K_s<$13.5 2MASS galaxies (as in Fig.~\ref{fig:2mass_mag}a) is compared to a linear theory 
$S$CDM prediction using input parameters of $\Omega_m$=1.0, $\Omega_b$=0.04, $h$=0.50 (dot-dashed line), a $\Lambda$CDM prediction using the Hubble 
Volume input parameters as before (dashed line), and the best fit power spectrum shape (for $l\le$30) of $\Gamma_{eff}$=0.14 (dotted line). 
In each case we use the best fit normalisation of ($\sigma_8b_K$)$^2$=1.36. The errorbars indicate the 1$\sigma$ spread determined from the 27 
mock 2MASS power spectra. In the lower panel we show the fractional deviation from the best fit $\Gamma_{eff}$=0.14 prediction.}
\label{fig:2mass_mod}
\end{center}  
\end{figure}

Our constraint on the $K_s$-band galaxy power spectrum normalisation of ($\sigma_8b_K$)$^2=1.36\pm0.10$ is also slightly higher than the \citet{mal} 
result of $\sigma_8b_K$=1.0$\pm$0.1. Using the WMAP-2dFGRS best fit matter power spectrum normalisation of $\sigma_8=0.84\pm0.04$ \citep{ben}, we 
constrain the $K_s$-band bias to $b_K=1.39\pm0.12$, in reasonable agreement with previous measurements determined from the 2MASS clustering dipole of 
$b_K=1.37\pm0.34$ \citep{mal2} and the 2MASS angular correlation function analysis of $b_K=1.1\pm0.1$ \citep{mal}. The constraint on the bias derived 
in this work rejects $b_K=1$ at $>3\sigma$; it appears therefore that galaxies selected in the $K_s$-band are clustered more strongly than both the 
underlying mass distribution and galaxies selected in optical wavebands for which $b\approx$1 \citep[e.g.][]{ver,gaz}.

We are also able to provide constraints on other cosmological parameters. We fit to $\Omega_m h$ and $\Omega_b$/$\Omega_m$ since these
primarily determine the shape of the input $P(k)$ and the size of the baryon oscillations. We determine model
angular power spectra in a 71$\times$51$\times$11 grid between 0.1$\le\Omega_m h\le$0.9, 0.0$\le\Omega_b$/$\Omega_m\le$0.5 and 0.4$\le h\le$0.9
(the effect of $h$ on the angular power spectrum is fairly small and we therefore use a lower resolution).
We perform least squares fits to the $K_s<$13.5, $|b|>$20$^{\circ}$ angular power spectrum at scales of $l\le$30, using errors determined for the 
2MASS datapoints as before which are independent of power spectrum normalisation.

The filled contours in Fig.~\ref{fig:conf} show the associated confidence regions marginalising over the normalisation. 
We are able to provide weak constraints on the cosmology of $\Omega_m h<$0.62 and $\Omega_b$/$\Omega_m<$0.46 (at 1$\sigma$ confidence). These 
constraints are particularly insensitive to the baryon density since the acoustic oscillations detected in redshift survey analyses are smoothed over 
a wide range of angular scales. However our constraints are in good agreement with the previous results at smaller scales from the 2dFGRS $P(k)$ 
\citep{per,cole2}. As an example of how our results can differentiate between different cosmological models we show the 2MASS galaxy angular power 
spectrum compared with $\Lambda$CDM and SCDM predictions in Fig.~\ref{fig:2mass_mod}.

We also wish to examine our assumption, used throughout this work, that the uncertainty due to cosmic variance determined from the 27 
$\Lambda$CDM mock 2MASS catalogues is independent of the power spectrum normalisation. To do this, we instead assume that the errors determined from 
the $\Lambda$CDM mock catalogues simply scale with the model power spectrum normalisation as would be the case in the cosmic variance limited 
scenario, and compare the two cases. In Figs.~\ref{fig:sigbb} and~\ref{fig:conf} we show the associated 1$\sigma$ confidence regions by the dashed 
lines, marginalising over the power spectrum normalisation. We find best fit parameters of $\Gamma_{eff}$=0.125$\pm$0.030, 
($\sigma_8 b_K$)$^2$=1.47$^{+0.27}_{-0.17}$, $\Omega_b/\Omega_m<$0.52 and $\Omega_mh<0.71$. This constraint on the galaxy power spectrum normalisation 
implies a $K_s$-band bias of $b_K$=1.44$^{+0.21}_{-0.14}$ (using the WMAP-2dFGRS constraint on $\sigma_8$ as before). It is clear that while 
the associated confidence regions for each parameter are slightly larger the results are in fair agreement whichever error analysis is used. However, 
it is clear from Fig.~\ref{fig:sigbb} that using this alternative assumption about the errors it is more difficult to reject combinations of high bias 
and steeper $\Gamma_{eff}$ slopes. For example, $\Gamma_{eff}$=0.05 would only be rejected at 2.5$\sigma$. More simulations of other cosmologies 
are needed to check whether these errors or the errors used elsewhere in this paper are most likely to be correct.

\section{Conclusions}
	
We have used 518~576 $K_s<13.5$, $|b|>20^{\circ}$ galaxies selected from the 2MASS full release extended source catalogue to determine the 
associated angular power spectrum and constrain the form of galaxy fluctuations to Gigaparsec scales. We have compared this to a $\Lambda$CDM N-body 
mock prediction constructed from the Hubble Volume simulation; it is in reasonable agreement although there is a discrepancy in the slopes at $l>$30 
in that the 2MASS result is significantly steeper than the mock prediction. We compare these to a linear theory prediction using the $\Lambda$CDM 
Hubble Volume simulation input parameters; there is good agreement with the mock prediction at scales where non-linear effects are insignificant 
($l\le$30).

Possible sources of systematic error were investigated. We first examined the effect of imposed magnitude limit; the 2MASS 
angular power spectrum slope was robust with respect to the 2MASS mock and model predictions. 
The 2MASS galaxy angular power spectrum is also robust to different magnitude estimators 
and zero-point corrections (imposed to agree with the \citet{lov} photometry) at the $\approx$10\% level. We correct for extinction using 
the \citet{shl} dust maps; the effect on the angular power spectrum is $\approx$10\% at $l$\lsim10, and $\approx$1\% at smaller scales. Our 
results are also robust to window function effects; the effect of a $|b|>20^{\circ}$ cut is \lsim5\% at all scales. 

Finally, we have used linear theory predictions for the 2MASS galaxy angular power spectrum formed from the transfer function 
fitting formulae of \citet{eis} to determine constraints on $\Omega_m h$ and $\Omega_b$/$\Omega_m$ assuming a flat CDM cosmology, a primordial 
$n_s$=1 spectrum and a negligible neutrino mass. Our results are in agreement 
with the 2dFGRS $P(k)$ constraints \citep{per}, and we are able to provide weak constraints of $\Omega_m h<$0.62 and 
$\Omega_b$/$\Omega_m<$0.46 (at 1$\sigma$ confidence). Angular power spectrum analysis is particularly insensitive to the baryon density since any 
associated baryon oscillations are likely to be smoothed over a wide range of angular scales. However, given the huge volume probed ($\approx$5 times 
the final 2dFGRS volume) the associated constraints on the power spectrum shape and normalisation are more significant. 
We also determine constraints for the galaxy power spectrum shape, $\Gamma_{eff}$, and normalisation, ($\sigma_8b_K$)$^2$. In agreement with the 
2dFGRS and WMAP values, we find that $\Gamma_{eff}=0.14\pm0.02$. This is slightly higher than an alternative value found by \citet{mal} using the 
2MASS dataset of $\Gamma_{eff}=0.116\pm0.009$, determined through an inversion of the angular correlation function. We also tightly 
constrain the $K_s$-band galaxy power spectrum normalisation to ($\sigma_8b_K$)$^2=1.36\pm0.10$. Using the WMAP-2dFGRS value of 
$\sigma_8=0.84\pm0.04$ \citep{ben}, this implies a $K_s$-band bias of $b_K=1.39\pm0.12$. 

We also investigated the likely impact on our assumption that the errors which we use to constrain various cosmological parameters, determined from 
the unbiased $\Lambda$CDM mocks, are independent of cosmology and power spectrum normalisation by instead assuming that these errors simply scale 
with the power spectrum normalisation as would be the case in the cosmic variance limited scenario. We find that while the associated confidence 
regions are slightly larger the results are in fair agreement. However it becomes less easy to reject models with lower $\Gamma_{eff}$; therefore 
although the data appears to prefer a $\Lambda$CDM power spectrum slope, it may still not be possible to rule out a ssignificantly steeper 
$\Gamma_{eff}$.

\section*{Acknowledgements} 

This publication makes use of data products from the 2 Micron All-Sky Survey, which is a joint project of the University of 
Massachusetts and the Infrared Processing and Analysis Centre/California Institute of Technology, funded by the Aeronautics and 
Space Administration and the National Science Foundation. We thank Shaun Cole for useful discussion, Tom Jarrett for his help with the 
2MASS magnitudes, Adrian Jenkins and Carlton Baugh for their assistance with the Hubble Volume mock catalogues, and Will Percival for 
providing the 2dFGRS constraints. We also thank David Johnston for useful comments.

\label{lastpage}

\end{document}